\documentclass[11pt,leqeq]{article}
\usepackage{amssymb,amsthm}
\usepackage{amsmath}
\usepackage{amscd}
\usepackage{xy}
\xyoption{all}
\usepackage[dvips]{graphicx}
\begin{document}

\setlength{\baselineskip}{16pt}

\title{An example of Feynman-Jackson integral}

\author{ Rafael D\'\i az and Eddy Parigu\'an}

\maketitle


\begin{abstract}
We review the construction of a $q$-analogue of the Gaussian
measure. We apply that construction to obtain a $q$-analogue of
Feynman integrals and to compute explicitly an example of such
integrals.
\end{abstract}

\section{Introduction}

{\noindent}The main goal of this note is to provide a gentle
introduction to the theory of Feynman-Jackson integrals, and to
compute an example of such integrals as explicitly as possible.
Roughly speaking Feynman-Jackson integrals are the analogue in
$q$-calculus of Feynman integrals. The name of Jackson is included
since integration in $q$-calculus was introduced by F.H. Jackson,
see \cite{Jac} and \cite{JA}. Our computations are done in the one
dimensional setting. Extending our concepts to higher dimensions,
in particular to infinite dimensions, is the main problem in this
subject. In principle such extension is possible but making the
notation easy to handle is hard, as the reader will learn by
looking at the simple case of one dimensional integrals.\\

{\noindent}The other fundamental task in this subject matter is to
find applications of Feynman-Jackson integrals in physics.
Although in this note we focus  on the mathematical properties of
such integrals, we believe that our formalism will find
applications thanks  to the following facts:

\begin{itemize}
\item In this paper our emphasis is on Feynman-Jackson integrals
which rely on the construction of a $q$-analogue of the Gaussian
measure. Given the widespread range of applications in mathematics
and physics alike of the Gaussian measure, we expect its
$q$-analogue to gradually find its natural set of applications.
The subject of $q$-probability theory and $q$-random processes is
still in a developing phase, but already a solid step forward has
been taken by Kupershmidt in \cite{kuper}.

\item{$q$-calculus is adapted to work with arbitrary functions
while the usual rules of calculus demand certain kind of
regularity. The $q$-Gaussian measure is likely to find
applications in the context of highly non-regular phenomena.}

\item{Jackson integrals are given by infinite sums. Cutting off the number of terms
appearing in such sums provides a natural regularization method
for diverging Jackson integrals.}

\item{Classically, we think of the continuous as a limit of the
discrete. Indeed, in the limit $q \rightarrow 1$ we recover from
Feynman-Jackson integrals the usual Feynman integrals. Thus our
Feynman-Jackson integrals provide a new method for computing
Feynman integrals: first compute the corresponding Feynman-Jackson
integral and then take the limit as $q$ goes to $1$.}

\item{From a quantum perspective, it is the continuous that should
be regarded as an approximation to the fundamental discrete
quantum theory. In particular it has been argued, see \cite{Ro},
\cite{sm}, that in the quantization of gravity discrete structures
will emerge naturally. If that is so, then integration over
discrete structures may become a fundamental issue. Our formalism
may shed some light as the form that such theory of discrete
integration may take. We emphasize that the calculus itself in
$q$-calculus is discrete while the variables involved remain
continuous. }
\end{itemize}

\section{Gauss-Jackson integrals}

{\noindent} Let us recall some notions of $q$-calculus, see
\cite{AND}, \cite{RA},\cite{Ch} and \cite{HY} for more
information. Fix a real number $0 < q < 1$. Let $f:\mathbb{R}
\rightarrow \mathbb{R}$ be a function and $x \in \mathbb{R}$ be a
real number. The $q$-derivative of $f$ at $x$ is given by
\begin{equation}\label{qD}
\partial_{q}f(x)=\frac{f(qx)-f(x)}{(q-1)x}.
\end{equation}
{\noindent}For example if $t \in \mathbb{R}$ then
$\partial_{q}x^{t} = [t]_{q}x^{t-1}$ where
$[t]_{q}=\frac{q^{t}-1}{q-1}$.\\

{\noindent}The $q$-integral, better known as the Jackson integral,
of $f$ between $0$ and $b \in \mathbb{R}^{+}$ of $f$ is given by
\begin{equation}\label{qI}
\int_{0}^{b}f(x)d_qx=(1-q)b\sum_{n=0}^{\infty}q^nf(q^nb).
\end{equation}

{\noindent}We also define

\begin{equation}\label{qI}
\int_{-b}^{0}f(x)d_qx=\int_{0}^{b}f(-x)d_qx
\end{equation}

{\noindent}and

\begin{equation}\label{qI}
\int_{-b}^{b}f(x)d_qx= \int_{-b}^{0}f(x)d_qx + \int_{0}^{b}f(x)d_qx.
\end{equation}

{\noindent} Notice that in the limit $q \rightarrow 1$ the
$q$-derivative and the $q$-integral approach the usual derivative
and the Riemann integral, respectively. The $q$-analogues of the
rules of derivation and integration by parts are
\begin{equation}\label{qdp}
\partial_{q}(fg)(x)=  \partial_{q}f(x)g(x) +
f(qx)\partial_{q}g(x),
\end{equation}

\begin{equation}\label{qIp}
\int_{0}^{b}\partial_{q}f(x)g(x)d_qx = -\int_{0}^{b}f(qx)\partial_{q}g(x)d_qx + f(b)g(b)-f(0)g(0).
\end{equation}

{\noindent}The first goal of this note is to describe the
$q$-analogue of the Gaussian measure on $\mathbb{R}$. The moments
of the Gaussian measure are given by the integrals

\begin{equation}\label{gm}
\frac{1}{\sqrt[]{2\pi}}\int_{-\infty}^{\infty}e^{-\frac{x^2}{2}}x^ndx.
\end{equation}

{\noindent}A remarkable property of the Gaussian measure is that
it provides a bridge between measure theory and combinatorics.
Indeed, the moments of the Gaussian measure are

\begin{equation}\label{gc1}
\frac{1}{\sqrt[]{2\pi}}\int_{-\infty}^{\infty}e^{-\frac{x^2}{2}}x^{2n}dx= (2n-1)(2n-3)...7.5.3.1,
\end{equation} and

\begin{equation}\label{gc2}
\frac{1}{\sqrt[]{2\pi}}\int_{-\infty}^{\infty}e^{-\frac{x^2}{2}}x^{2n+1}dx= 0.
\end{equation}
{\noindent}The number $(2n-1)(2n-3)...7.5.3$ is often denoted by
$(2n-1)!!$ and is called the double factorial. The reader may
consult \cite{ED} for a natural generalization of such numbers. It
can be shown that $(2n-1)(2n-3)...7.5.3$ counts the number of
pairings on the set $[2n]=\{1,2,...,2n\}.$ A pairing on $[2n]$ is
a partition of $[2n]$ into $n$ blocks each of cardinality two. So
for example we have that

\begin{equation}\label{gcex}
\frac{1}{\sqrt[]{2\pi}}\int_{-\infty}^{\infty}e^{-\frac{x^2}{2}}x^{4}dx= 3,
\end{equation}

\begin{figure}[h!]
\begin{center}
\includegraphics[width=5.5in]{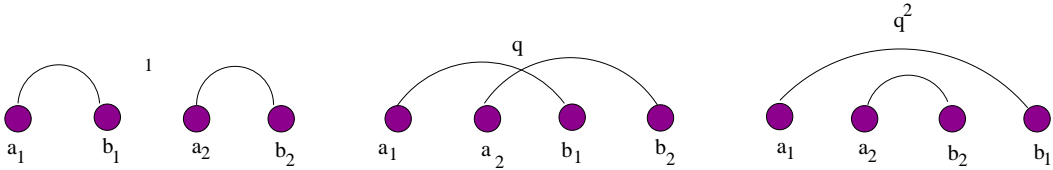}
\caption{Combinatorial interpretation of $[3]_q.$
\label{fig:Pochha graphs}}
\end{center}
\end{figure}

{\noindent} since as shown in Figure \ref{fig:Pochha graphs} there
are $3$ pairings on a set with $4$ elements.

{\noindent}Thus we  see that the Gaussian measure has a clear cut
combinatorial meaning. This simple fact explains the source of
graphs in the computation of Feynman integrals. In order to define
the $q$-analogue of the Gaussian measure we must find
$q$-analogues for the objects appearing in the Gaussian measure,
namely $\sqrt[]{2\pi},\infty, e^{-\frac{x^2}{2}}, x^n$ and $dx$.
The Lebesgue measure $dx$ agrees with Riemann integration for good
functions. Thus it is only natural to replace $dx$ by Jackson
integration $d_{q}x.$ While the factor $x^{n}$ remains unchanged,
finding the $q$-analogue of $e^{-\frac{x^2}{2}}$ is actually quite
a subtle matter. First, we must find a $q$-analogue for the
exponential function $e^x$ which is characterized by the
properties $\partial e^x =e^x$ and $e^0 =1$. So we look for a
function $e_{q}^x$ such that  $\partial e_{q}^x =e_{q}^x$ and
$e_{q}^0 =1.$ A solution to this couple of equations is
\begin{equation}\label{qex1}
e_{q}^x = \sum_{n=0}^{\infty}\frac{x^n}{[n]_{q}!},
\end{equation} where
$$[n]_{q}!=[n]_{q}[n-1]_{q}[n-2]_{q}...[2]_q
\mbox{ \ \ and \ \ }
[n]_{q}= 1 + q + q^2 + q^3 + ... + q^{n-1}.$$

{\noindent}The $q$-analogue of the identity $e^{x}e^{-x}=1$ is
$e_{q}^{x}E_{q}^{-x}=1$, where

\begin{equation}\label{qex2}
E_{q}^{x}=
\sum_{n=0}^{\infty}q^{\frac{n(n-1)}{2}}\frac{x^n}{[n]_{q}!}.
\end{equation}
{\noindent}Once we have obtained $q$-analogues for the exponential
map and its inverse one might think that it is straightforward to
generalized the term $e^{-\frac{x^2}{2}}$ of the Gaussian
integrals. However, this is not the case and while our first
impulse is to try $E_{q}^{-\frac{x^2}{[2]_q}}$, the right answer
\cite{CTT} is  to replace $e^{-\frac{x^2}{2}}$ by

\begin{equation}\label{qex3}
E_{q^{2}}^{-\frac{q^{2}x^2}{[2]_q}}=
\sum_{n=0}^{\infty}\frac{(-1)^{n}q^{n(n+1)}x^{2n}}{(1+q)^{n}[n]_{q^{2}}!}.
\end{equation}
{\noindent}Now we consider the integration limits. It is amusing
that whereas the classical Gaussian measure is given by an
improper integral, its $q$-analogue turns out to be a definite
integral whose limits depend on $q$ and go to (plus or minus)
infinity as $q$ approaches to $1$. A similar situation occurs with
the integral representations  of the $q$-analogue of the gamma
function, see \cite{K} and \cite{So}. Indeed, without further
motivation we shall take the boundary limits in the Gaussian
integrals to be $-\nu$ and $\nu$ where $\nu=\frac{1}{\sqrt{1-q}}.$
To find the $q$-analogue $c(q)$ of the $\sqrt{2\pi}$\ appearing in
Gaussian integrals we must demand that $c(q)$ be such that
\begin{equation}\label{qG}
\frac{1}{c(q)}\int_{-\nu}^{\nu}
E_{q^{2}}^{\frac{-q^2x^2}{[2]_q}}d_qx = 1.
\end{equation}
{\noindent} Thus $c(q)$ is given by
\begin{equation}\label{c(q)1}
c(q) = \int_{-\nu}^{\nu} E_{q^2}^{\frac{-q^2x^2}{[2]_q}}d_qx =
2\int_{0}^{\nu} E_{q^{2}}^{\frac{-q^2x^2}{[2]_q}}d_qx =
2(1-q)\nu\sum_{n=0}^{\infty}q^n
E_{q^{2}}^{\frac{-q^2(q^n\nu)^2}{[2]_q}},
\end{equation} so

\begin{equation}\label{c(q)2}
c(q) = 2\sqrt{1-q}\sum_{n=0}^{\infty}\sum_{m=0}^{\infty}
\frac{(-1)^{m}q^{m(m+1)+(2m+1)n}}{(1-q^2)^{m}[m]_{q^{2}}!}
\end{equation}

{\noindent}and interchanging the order of summation we get the
identity

\begin{equation}\label{c(q)2}
c(q) = 2\sqrt{1-q}\sum_{m=0}^{\infty}
\frac{(-1)^{m}q^{m(m+1)}}{(1-q^{2m+1})(1-q^2)^{m}[m]_{q^{2}}!}.
\end{equation} Since $\mbox{Lim}_{q \rightarrow 1}c(q)=\sqrt{2\pi}$ we obtain the
amusing identity

\begin{equation}\label{pi}
\sqrt{2\pi} = 2  \mbox{Lim}_{q \rightarrow 1}
\sqrt{1-q}\sum_{m=0}^{\infty}
\frac{(-1)^{m}q^{m(m+1)}}{(1-q^{2m+1})(1-q^2)^{m}[m]_{q^{2}}!}.
\end{equation}

{\noindent} We now look for the $q$-analogue of identities
$(\ref{gc1})$ and $(\ref{gc2})$ for the moments of Gaussian
integrals. A key result is that one can show the identities

\begin{equation}\label{qgc1}
\frac{1}{c(q)}\int_{-\nu}^{\nu}E_{q^{2}}^{\frac{-q^2x^2}{[2]_q}}x^{2n}d_{q}x=
[2n-1]_{q}[2n-3]_{q}...[7]_{q}[5]_{q}[3]_{q}[1]_q=[2n-1]_q!!,
\end{equation} and

\begin{equation}\label{qgc2}
\frac{1}{c(q)}\int_{-\nu}^{\nu}E_{q^{2}}^{\frac{-q^2x^2}{[2]_q}}x^{2n+1}d_{q}x= 0.
\end{equation}

{\noindent} Identity $(\ref{qgc2})$ follows from the fact that
$x^{2n+1}$ is an odd function and
$E_{q^{2}}^{\frac{-q^2x^2}{[2]_q}}$ is an even function. Identity
$(\ref{qgc1})$ is proved recursively. Using formula $(\ref{qIp})$,
one shows that
\begin{equation}\label{qgc3}
\frac{1}{c(q)}\int_{-\nu}^{\nu}E_{q^{2}}^{\frac{-q^2x^2}{[2]_q}}x^{2n+2}d_{q}x=
\frac{[2n+1]_{q}}{c(q)}\int_{-\nu}^{\nu}E_{q^{2}}^{\frac{-q^2x^2}{[2]_q}}x^{2n}d_{q}x.
\end{equation}
{\noindent} Identities (\ref{qG}) and (\ref{qgc3}) imply identity
(\ref{qgc1}).  Next we describe a combinatorial interpretation of
the number
\begin{equation}\label{qf}
[2n-1]_q!!=[2n-1]_{q}[2n-3]_{q}...[7]_{q}[5]_{q}[3]_{q}[1]_q.
\end{equation}

{\noindent}An ordered pairing $p$ on $[2n]=\{1,2,...,2n\}$  is a
sequence $p=\{
\ (a_i,b_i) \
\}_{i=1}^{n} \in ([2n]^2)^n$ such that
\begin{itemize}
\item {$a_1< a_2 < \dots < a_n$.}
\item{ $a_i<b_i,\quad
i=1,\dots,n.$}
\item{ $\displaystyle{[2n]=\bigsqcup_{i=1}^{n} \{a_i,b_i\}.}$}
\end{itemize}

{\noindent} We denote by $P[2n]$ the set of ordered pairings on
$[2n]$. We are going to define a weight $w(p)$ for each $p \in
P[2n]$. Let us introduce the following notation
\begin{itemize}
\item {$((a_i,b_i))=\{j\in [[2n]]: a_i<j<b_i\}$ for all $(a_i,b_i)\in
p$.}
\item {$B_i(p)=\{b_j: 1 \leq j < i \}$.}
\end{itemize}
{\noindent} The weight of $p$ is defined by the rule

\begin{equation}\label{peso}
\displaystyle{w(p)=\prod_{i=1}^{n}q^{|((a_i,b_i)) \setminus
B_i(p)|}=q^{\sum_{i=1}^{n}|((a_i,b_i)) \setminus B_i(p)|}}.
\end{equation}

{\noindent} Figure \ref{fig:1} and Figure \ref{fig:2} below show a
couple of examples of pairings together with the corresponding
weights

\begin{figure}[h!]
\begin{center}
\includegraphics[width=3in]{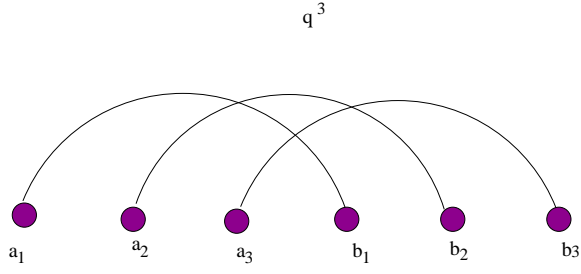}
\caption{ A Pairing with weight $q^3$.\label{fig:1}}
\end{center}
\end{figure}

\begin{figure}[h!]
\begin{center}
\includegraphics[width=3in]{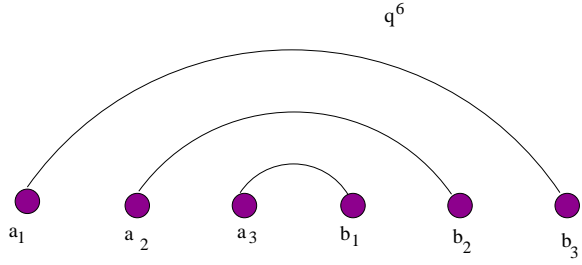}
\caption{A Pairing with weight $q^6$.\label{fig:2}}
\end{center}
\end{figure}
\newpage
{\noindent} Using this language we can state the following result
which is proved by induction

\begin{equation}\label{intpocha}
[2n-1]_q!!= \sum_{p \in P[2n]}w(p).
\end{equation}

{\noindent} Notice that as $q\longrightarrow 1$ we recover from
$(\ref{intpocha})$  the well-known identity
$$(2n-1)!!=|\{\mbox{pairings on}\ [[2n]]\}|.$$

{\noindent} For example $[3]_q = 1+q+q^{2}$ which agrees with the
sum of the weights of the three pairings on $[4]$ shown in Figure
$\ref{fig:Pochha graphs}$.

{\noindent} In conclusion we have proved that
\begin{equation}\label{qgc5}
\frac{1}{c(q)}\int_{-\nu}^{\nu}E_{q^{2}}^{\frac{-q^2x^2}{[2]_q}}x^{2n}d_{q}x=
\sum_{p \in P[2n]}w(p).
\end{equation}

\section{Examples of Feynman-Jackson integrals}

{\noindent} We want to study the computation of Feynman-Jackson
integrals of the form
\begin{equation}\label{int24}
I(g)=\frac{1}{c(q)}\displaystyle{\int_{-\nu}^{\nu}
E_{q^2}^{\frac{-q^2x^2}{[2]_q}+g \frac{x^3}{[3]_q!}}} d_qx.
\end{equation}

{\noindent}As usual in the theory of Feynman integrals we regard
$I(g)$ as a formal power series in the formal variable $g$, i.e.,
$I(g)\in\mathbb{C}[[g]]$. The first step in the computation of a
Feynman-Jackson integrals is to reduce it to the computation of a
countable number of Gaussian-Jackson integrals. This step is carry
out with the help of following formula proved in \cite{ED1}

\begin{equation}
E_{q^2}^{x+y}= E_{q^2}^{x}\left( \sum_{c,d\geq 0}
\lambda_{c,d}x^{c}y^{d}\right)
\end{equation}

{\noindent} where $\lambda_{c,d}={\displaystyle\sum_{k=0}^c
\frac{(-1)^{c-d}{d+k\choose k}q^{(d+k)(d+k-1)}
}{[d+k]_{q^{2}}![c-k]_{q^{2}}!}}$. Making the substitutions
${\displaystyle x\rightarrow -\frac{q^{2}x^{2}}{[2]_q}}$ and \\
${\displaystyle y\rightarrow g\frac{x^{3}}{[3]_q!}}$ we obtain

\begin{equation} {\displaystyle
E_{q^2}^{-\frac{q^{2}x^{2}}{[2]_q}+g\frac{x^{3}}{[3]_q!}}=
E_{q^2}^{-\frac{q^{2}x^{2}}{[2]_q}}\sum_{c,d\geq 0} \lambda_{c,d}
\frac{(-1)^{c}q^{2c} x^{2c} x^{3d}
}{[2]_q^{c}([3]_q!)^{d}}g^{d}=E_{q^2}^{-\frac{q^{2}x^{2}}{[2]_q}}\sum_{c,d\geq
0} \lambda_{c,d} \frac{(-1)^{c}q^{2c} x^{2c+3d}
}{[2]_q^{c}([3]_q!)^{d}}g^{d} },
\end{equation}
{\noindent} so we get

\begin{equation}\label{26}
{\displaystyle
E_{q^2}^{-\frac{q^{2}x^{2}}{[2]_q}+g\frac{x^{3}}{[3]_q!}}=\sum_{c,d,k}
\frac{(-1)^{2c-k} {d+k\choose k}
q^{(d+k)(d+k-1)+2c}}{[2]_q^c([3]_q!)^{d}[d+k]_{q^{2}}![c-k]_{q^{2}}!}
x^{2c+3d} g^d }.
\end{equation}

{\noindent} If we $q$-integrate both sides of equation (\ref{26})
we get

\begin{eqnarray}
{\displaystyle \frac{1}{c(q)}
\int_{-v}^vE_{q^2}^{-\frac{q^{2}x^{2}}{[2]_q}+g\frac{x^{3}}{[3]_q!}}}
d_qx&=&{\displaystyle\sum_{c,d,k} \frac{(-1)^{2c-k} {2d+k\choose
k} q^{(2d+k)(2d+k-1)+2c}
[2c+6d-1]_q!!}{[2]_q^c([3]_q!)^{2d}[d+k]_{q^2}! [c-k]_{q^2}!}g^{2d}} \\
\mbox{ }& \mbox{ } & \mbox{ }\nonumber \\
&=&{\displaystyle \sum_{c,d,k} \frac{(-1)^{2c-k} {2d+k\choose k}
q^{(2d+k)(2d+k-1)+2c}
[2c+6d-1]_q!!}{[2]_q^c([3]_q!)^{2d}[2d+k]_{q^2}!
[c-k]_{q^2}!}g^{2d}}\label{e2}
\end{eqnarray}

{\noindent} The second step in the computation of a Feynman
integral is to write such as integral as a sum of a countable
number of contributions, where each summand is naturally
associated to certain kind of graph, see \cite{ET}. So we want to
understand the right hand side of the equation (\ref{e2}) in terms
of a summation of the weights of an appropriated set of
isomorphism classes of graphs. Consider the category
$\mathbf{Graph}_q^3$ whose objects are planar graphs $(V,E,b)$
such that

\begin{enumerate}
\item{$V=\{ \bullet\}\sqcup V_1\sqcup V_2$, where
$V^1=\{\otimes_1,\dots,\otimes_{c}\}$ and
$V^2=\{\circ_1,\dots,\circ_{d}\}$.} \item{$E=E_1\sqcup E_2\sqcup
E_3$.} \item{$b:E \longrightarrow P_2(V)=\{A\subset V|\ 1\leq
|A|\leq2\}$.}
\end{enumerate}

{\noindent}We use the following notation
$$I(v,e)= \left\{
\begin{array}{ll}
  0, & \mbox{if}\ v\not\in b(e)  \\
  1, & \mbox{if}\ |b(e)|=2, v\in b(e) \\
  2, & \mbox{if}\ |b(e)|=\{v\} \\
\end{array}\right.$$

{\noindent}This data must satisfy the following axioms:

\begin{enumerate}
\item{${\displaystyle \sum_{e\in E_3}I(\otimes_i,e)=2}$ and
${\displaystyle \sum_{e\in E_3}}I(\circ_j,e)=3$. }
\item{$|b^{-1}(\otimes_i,\bullet)|\leq 1$ and if
$|b^{-1}(\otimes_i,\bullet)|=1$ then
$|b^{-1}(\otimes_j,\bullet)|=1$ for $i\leq j\leq c$.} \item{If
$e\in E_1\sqcup E_2$, then $b(e)=\{\otimes_i,\bullet\}$ or
$b(e)=\{\circ_j,\circ\}$ for some  $1\leq i\leq c$ or$1\leq j\leq
d$. }
\item{If $e\in E_3$ then $\bullet\not\in b(e)$.}
\item{$|E_2|\leq |V_1|$.}

{\noindent} To each graph $\Gamma=(V,E,b)$ as above we associate
two polynomials in $q$, $\omega_q(\Gamma)$ and $a_q(\Gamma)$,
given respectively, by
\begin{enumerate}
\item{$\omega_q(\Gamma)=(-1)^{|E_1|}
q^{2|V_1|+2{|V_2|+|E_2|\choose 2}}\omega(p)$. Above $p$ is the
natural pairing induced by $E_3$ on the flags of $\Gamma$,
i.e.,the set $\{ (\otimes_i,e)|\  \otimes_i\in e\}\sqcup
\{(\circ_j,e)|\
\circ_i\in e\}$. $\omega(p)$ is the weight of $p$ as given by
formula (\ref{peso}).}
\item{$a_q(\Gamma)=[2]_q^{|V_1|}([3]_q!)^{|V_2|}[|V_2|+|E_2|]_{q^2}![|V_1|-|E_2|]_{q^2}!$.}

\end{enumerate}

\end{enumerate}

\begin{figure}[h!]
\begin{center}
\includegraphics[width=2in]{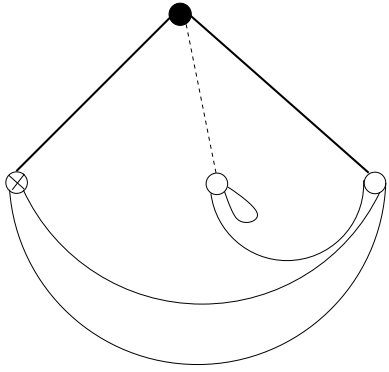}
\caption{Feynman graph $\Gamma$ with $\omega_q(\Gamma)=-q^{16}$
and $ a_q(\Gamma)=[2]_q^4[3]_q^3$.\label{fig:wq}}
\end{center}
\end{figure}

\begin{figure}[h!]
\begin{center}
\includegraphics[width=2in]{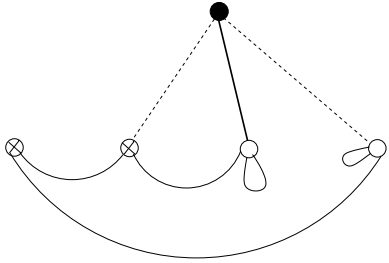}
\caption{Feynman graph $\Gamma$ with $\omega_q(\Gamma)=q^{22}$ and
$ a_q(\Gamma)=[2]_q^2([3]_q!)^2[4]_q!$. \label{fig:wq1}}
\end{center}
\end{figure}

{\noindent} Figure \ref{fig:wq} and Figure \ref{fig:wq1} show
examples of graphs in $\mathbf{Graph}_q^3$  together with the
corresponding polynomials $a_q$ and $\omega_q$.

\newpage

{\noindent} Now we are ready to state the main result of this
paper, namely, we write the integral (\ref{int24}) as a sum over
the weights of graphs as follows

\begin{equation}\label{ult}
{\displaystyle {\displaystyle \frac{1}{c(q)}
\int_{-v}^vE_{q^2}^{-\frac{q^{2}x^{2}}{[2]_q}+g\frac{x^{3}}{[3]_q!}}}
d_qx=\sum_{(V,E,b)\in{\mathbf{Graph}_q^3}/\sim }
\frac{\omega_q(\Gamma)}{a_q(\Gamma)}g^{|V_2|} }
\end{equation}
{\noindent} where the sum runs over all isomorphisms classes of
graphs in $\mathbf{Graph}_q^3$. Identity (\ref{ult}) follows
directly from identity (\ref{e2}) and the definitions above.

\section*{Acknowledgment}

We thank Bernardo Uribe  and the math department at Uniandes where
this work was finished.

$$\begin{array}{c}
  \mbox{Rafael D\'\i az. Universidad Central de Venezuela (UCV).} \ \  \mbox{\texttt{ragadiaz@gmail.com}} \\
\mbox{Eddy Parigu\'an. Universidad de los Andes. Bogota. (Uniandes).} \ \  \mbox{\texttt{eparigua@uniandes.edu.co}} \\
\end{array}$$


\begin{thebibliography}{10}

\bibitem{AND}
G.~Andrews, \emph{The theory of partitions}, Encyclopedia of
Mathematics and
  its Applications, 1938.

\bibitem{RA}
G.~Andrews, R.~Askey, and R.~Roy, \emph{Special functions},
Cambridge
  University Press, 1999.

\bibitem{ED}
Rafael~D\'\i az and Eddy Pariguan, \emph{On hypergeometric
functions and
  {Pochhammer} $k$-symbol}, to appear in Divulgaciones
  Matem\'aticas,  math.CA/0405596.

\bibitem{ED1}
Rafael~D\'\i az and Eddy Pariguan, \emph{Feynman-Jackson
integrals}, Journal of Nonlinear Mathematical Physics 13 (2006),
  no.~3, 365--376.
\bibitem{CTT}
Rafael~D\'\i az and Carolina Teruel, \emph{q,k-generalized gamma
and beta
  functions}, Journal of Nonlinear Mathematical Physics 12 (2005),
  no.~1, 118--134.

\bibitem{Ch}
P.~Cheung and V.~Kac, \emph{Quantum calculus}, Springer-Verlag,
2002.

\bibitem{ET}
Pavel Etingof, \emph{Mathematical ideas and notions of quantum
field theory},
  Preprint.

\bibitem{HY}
G.~Gasper and M.~Rahman, \emph{Basic hypergeometric series},
Cambridge
  University Press, 1990.

\bibitem{Jac}
F.H. Jackson, \emph{A generalization of the functions $\gamma(n)$
and $x^n$},
  Porc. Roy Soc. London 74 (1904), 64--72.

\bibitem{JA}
F.H. Jackson, \emph{On q-definite integrals}, Quart. J. Pure Appl.
Math. 41
  (1910).

\bibitem{K}
H.T. Koelink and Koornwinder, \emph{q-special functions, in
deformation theory
  and quantum groups with applications to mathematical physics}, Edited by
  {M}urray {G}erstenhaber and {J}im {S}tasheff, Amer. Math Soc 134
  (1992), 141--142.

\bibitem{kuper}
Boris Kupershmidt, \emph{q-Probability: I. Basic Discrete
Distributions}, Journal of Nonlinear Mathematical Physics 7
(2000), no.~1, 73--93.


\bibitem{Ro}
Rovelli Carlo and Smolin Lee, \emph{Discreteness of area and
volume in quantum gravity}, Nuclear Phys, B442, 3, (1995),
593-619.
\bibitem{sm}
Smolin, Lee, \emph{The physics of spin networks}, The geometric
universe, Oxford Univ. Press, Oxford, (1998), 299-304.
\bibitem{So}
A.~De Sole and V.~Kac, \emph{On integral representations of
q-gamma and q-beta
  functions},Atti. Accad. Naz. Lincei Cl. Sci. Fis. Mat. Natur. Rend. Lincei
(9). Mat. Appl. 16, 41, (2005), 11-29.

\end{thebibliography}
\end{document}